\documentclass[12pt,a4paper]{iopart}
\usepackage{graphicx,cite}
\usepackage{bm,bbm}
\usepackage{iopams}

\begin{document}

\renewcommand{\vec}[1]{\boldsymbol{#1}}
\newcommand{\qq}{\vec{q}}
\newcommand{\pp}{\vec{p}}
\newcommand{\abs}[1]{\left| #1 \right|}
\newcommand{\ee}{\mathrm{e}}
\newcommand{\dd}{\mathrm{d}}
\newcommand{\ii}{\mathrm{i}}
\renewcommand{\vec}[1]{\boldsymbol{#1}}
\newcommand{\rr}{\vec{r}}
\newcommand{\ff}{\vec{f}}
\newcommand{\generator}{\vec{g}}
\newcommand{\xx}{\vec{x}}
\newcommand{\yy}{\vec{y}}
\newcommand{\zz}{\vec{z}}
\newcommand{\mm}{\vec{m}}
\newcommand{\RR}{\vec{R}}
\newcommand{\eps}{\varepsilon}
\newcommand{\p}{\partial}
\newcommand{\ad}[1]{\mathrm{ad}_{#1}}
\newcommand{\order}[1]{\mathcal{O}\!\left({#1}\right)}
\newcommand{\Lie}[2]{\left[#1,#2\right]}
\newcommand{\Dop}[1]{\mathcal{D}_{\!#1}}
\newcommand{\Lop}[1]{\mathcal{L}_{\!#1}}
\newcommand{\strich}[1]{\left.#1\right|}
\newcommand{\Htilde}{\tilde{H}}
\newcommand{\Hhat}{\hat{H}}
\newcommand{\Hdensity}{\mathcal{H}}
\newcommand{\AZ}[1]{``#1''}
\newcommand{\nmax}{{n_\mathrm{max}}}

\title[TST for wave packet dynamics. I. Thermal decay in metastable Schr\"odinger systems]{Transition state theory for wave packet dynamics. \\ I. Thermal decay in metastable Schr\"odinger systems}

\author{Andrej Junginger, J\"org Main, G\"unter Wunner, and Markus Dorwarth}

\address{1. Institut f\"{u}r Theoretische Physik, Universit\"{a}t Stuttgart, 70550 Stuttgart, Germany}

\date{\today}


\begin{abstract}
We demonstrate the application of transition state theory to wave packet dynamics in metastable Schr\"odinger systems which are approached by means of a variational ansatz for the wave function and whose dynamics is described within the framework of a time-dependent variational principle. The application of classical transition state theory, which requires knowledge of a classical Hamilton function, is made possible by mapping the variational parameters to classical phase space coordinates and constructing an appropriate Hamiltonian in action variables. This mapping, which is performed by a normal form expansion of the equations of motion and an additional adaptation to the energy functional, as well as the requirements to the variational ansatz are discussed in detail. The applicability of the procedure is demonstrated for a cubic model potential for which we calculate thermal decay rates of a frozen Gaussian wave function. The decay rate obtained with a narrow trial wave function agrees perfectly with the results using the classical normal form of the corresponding point particle. The results with a broader trial wave function go even beyond the classical approach, i.e., they agree with those using the quantum normal form. The method presented here will be applied to Bose-Einstein condensates in the following paper [A. Junginger, M. Dorwarth, J. Main, and G. Wunner, following paper, submitted to J. Phys. A].
\end{abstract}

\maketitle

\section{Introduction}
Transition state theory (TST) is one of the most important tools for analysing chemical reactions qualitatively as well as quantitatively and has also applications remote from its origin, e.g. in atomic and semiconductor physics \cite{Jaffe1999, Eckhardt1995}, the study of clusters \cite{Komatsuzaki1999}, and cosmology \cite{Oliveira2002}.

Fundamental to all these calculations is a classical Hamilton function $H(\qq,\pp)$ given in phase space coordinates $\qq,\pp$ which is either an inherent part of the underlying model or, e.g. for chemical reactions, given by an ab initio calculated potential energy surface $V(\qq)$ \cite{Garrett1983, Aoiz1994, Truong1990, Soto1992}.

In this paper and the following \cite{Junginger2011c}, we apply TST to wave packets, which are widely used for variational approaches to quantum mechanical systems, such as Bose-Einstein condensates \cite{Rau2010a, Rau2010b}. Here, the Schr\"odinger or Gross-Pitaevskii equation is solved within the Hilbert subspace of the variational parameters, and to describe the dynamics of the wave function, one makes use of a time-dependent variational principle (TDVP). 

In order to apply TST to a system with a known Hamilton function, one transforms the latter to its normal form using canonical transformations \cite{Waalkens2008}. In contrast, for the application of TST to wave packets, where such a Hamiltonian is \emph{not} known, we demonstrate a method to locally map the variational parameters to action variables in classical phase space using non-canonical transformations. This step is performed by a normal form expansion \cite{Murdock2010} of the equations of motion derived from the TDVP and the corresponding transformation of the energy functional. After having expanded the equations of motion into normal form, they are integrated to a common Hamilton function and an additional transformation guarantees its equivalence to the energy functional, which at the end serves as the classical Hamiltonian.

To demonstrate the procedure, we apply this method to calculating the decay rate of a frozen Gaussian wave function located at the metastable ground state of a cubic model potential and compare it with the ones resulting from ordinary classical and quantum normal forms \cite{Waalkens2008} of the corresponding point particle. For the application to Bose-Einstein condensates with \emph{coupled} Gaussian wave functions, we refer the reader to reference \cite{Junginger2011c}.

Our paper is organized as follows: First, we give a brief review of the classical and quantum normal form expansions, which can be applied to a given Hamiltonian $H(\qq,\pp)$. Then, we introduce the variational approach, demonstrate the general concept of mapping the variational parameters to classical phase space and illustrate the calculation of the respective thermal decay rates. At the end, we apply the three methods to the model potential and compare their results.

\section{Theory}

\subsection{Classical and quantum normal form}

We begin by summarizing the classical and quantum normal form transformation of a given Hamilton function
\begin{equation}
	H(\qq,\pp) = T(\pp) + V(\qq)
	\label{eq-original-Hamiltonian}
\end{equation}
in the vicinity of an equilibrium point $\qq_0$. Since this is a well known theory, we will only give an overview of the main aspects and refer the reader to Waalkens \etal\ \cite{Waalkens2008} and references therein for details.

In order to obtain the normal form Hamiltonian, one expands the Hamilton function \eref{eq-original-Hamiltonian} into its power series around the fixed point $\qq_0$, 
\begin{equation}
	H = \sum_{n=0}^\infty H_n.
\end{equation}
Here, $H_n$ summarizes monomials which are homogeneous of degree $n$, and for simplicity, we omit the dependency of the phase space variables $\qq,\pp$ from now on.

After having shifted the fixed point to the origin and having applied a symplectic transformation which \AZ{simplifies} the quadratic part $H_2$ in a way that diagonalizes the corresponding Hamiltonian equations of motions, the Hamilton function is in classical normal form with respect to the quadratic part. The aim is to transform the whole Hamilton function $H$ into normal form, which is defined by the condition \cite{Waalkens2008}
\begin{equation}
	\ad{H_2} H = \{ H_2, H \} = 0
\end{equation}
and where the adjoint operator $\ad{H_2}H = \{ H_2, H \}$ is defined by the Poisson bracket.

To transform the higher order terms $n \geq 3$, we assume the Hamiltonian to be in normal form up to order $n-1$ which we denote by the superscript $H^{(n-1)}$ and apply successive symplectic transformations given by the generator $W_n$. This generator $W_n$, which is a solution of the homological equation
\begin{equation}
	H_n^{(n)} = H_n^{(n-1)} - \left\lbrace H_2^{(2)} , W_n \right\rbrace,
\end{equation}
then transforms the Hamilton function according to 
\begin{equation}
	H^{(n)} = \sum_{k=0}^\infty \frac{1}{k!} [\mathrm{ad}_{W_n} ]^k H^{(n-1)}.
	\label{HCNF}
\end{equation}	
These steps are performed successively for each order $n$ until the Hamiltonian is in classical normal form $H_\mathrm{CNF}$ up to the desired order.

In analogy to the classical version a quantum normal form can be calculated by replacing the adjoint operator $\ad{W}$ by the Moyal adjoint action \cite{Waalkens2008, Schubert2006}
\begin{equation}
	\mathrm{Mad}_W H = \frac{2}{\hbar} W \sin \left( \frac{\hbar}{2} \left[ \left< \overleftarrow{\p}_{\!\!p} , \overrightarrow{\p}_{\!\!q} \right> - \left< \overleftarrow{\p}_{\!\!q} , \overrightarrow{\p}_{\!\!p} \right> \right] \right) H.
\end{equation}
Since the procedure of getting the quantum normal form is very similar to that of its classical analogue, we skip the details here and again refer the reader to references \cite{Waalkens2008, Schubert2006}.

Altogether, for a one degree of freedom Hamilton function of the form
\begin{equation}
	H = \frac{p^2}{2} + V(q),
	\label{eq-Ham-one-degree-of-freedom}
\end{equation}
with stationary point $q_0$ as we will use it for comparisons with the formalism for wave packets in section \ref{sec-results}, the quantum normal form Hamiltonian $H_\mathrm{QNF}$ can explicitly be written down in terms of the action variable $J$, and in order $\order{J^3}$ its symbol reads \cite{Waalkens2008}
\begin{eqnarray}
\fl
	H_\mathrm{QNF}^{(6)} &=
	V(q_0) + \lambda J + \frac{J^2}{16\lambda^2} \left( V^{(4)} + \frac{5 \left[ V''' \right]^2}{3\lambda^2} \right) - \frac{\hbar^2}{64 \lambda^2} \left( V^{(4)} + \frac{7 \left[ V''' \right]^2}{9\lambda^2} \right)  \nonumber \\
\fl	&- \frac{J^3}{288\lambda^3} \left( \frac{75 \left[ V''' \right]^2 V^{(4)}}{4\lambda^4} + \frac{17 \left[ V^{(4)}\right]^2 }{8\lambda^2} + \frac{235 \left[ V''' \right]^4 }{24\lambda^6} + \frac{7  V'''  V^{(5)}}{\lambda^2} + V^{(6)} \right) \nonumber \\
\fl	&+ \frac{5 \hbar^2 J}{1152 \lambda^3} \left( \frac{153 \left[ V''' \right]^2 V^{(4)}}{20\lambda^4} + \frac{67 \left[ V^{(4)}\right]^2 }{40\lambda^2} + \frac{77 \left[ V''' \right]^4 }{24\lambda^6} + \frac{19  V'''  V^{(5)}}{5\lambda^2} + V^{(6)} \right).
	\label{eq-normal-form-6}
\end{eqnarray}
Here $\lambda = \sqrt{-V''(0)}$ is the eigenvalue of the linearised Hamiltonian equations of motion and all derivatives are evaluated at $q=q_0$. The classical normal form $H_\mathrm{CNF}$ of the Hamiltonian \eref{eq-Ham-one-degree-of-freedom} can easily be obtained from equation \eref{eq-normal-form-6} by setting $\hbar \to 0$.

\subsection{Variational approach}

We now focus on the system from a different point of view and describe it quantum mechanically via its wave function $\psi(\rr,t)$. Its physical properties are given by the Hamilton operator $\Hhat$ and the time evolution of the wave function is determined by the time-dependent Schr\"odinger equation
\begin{equation}
	\Hhat \psi(\rr,t) = \ii \p_t \psi(\rr,t) ,
	\label{eq-Schroedinger-equation}
\end{equation}
where we set $\hbar=1$ from now on.

In the following, we will approach the system in the framework of a variational ansatz
\begin{equation}
	\psi(\rr,t) = \psi(\rr,\zz(t))
	\label{eq-wave-packet}
	\label{eq-wave-function-parameter}
\end{equation}
for the wave function in the form of a wave packet. Here $\zz(t)=(z_1, \ldots,z_d)^T$ is a set of $d$ complex and time-dependent variational parameters and the time evolution of the wave function is fully determined by that of the parameters. 

A usual way of treating such a variational ansatz is by means of a variational principle \cite{McLachlan1964} applied to the time-dependent Schr\"odinger equation \eref{eq-Schroedinger-equation}. However, we will also discuss it from the perspective of field theory with the Hamiltonian density $\Hdensity$. Although the latter approach will give exactly the same equations of motion for the variational parameters and the same energy functional, it allows for a better insight into the ``Hamiltonian relation'' between those two, and is the basic foundation of the variational approach and the subsequently performed normal form expansion.

\subsubsection{The time-dependent variational principle}

An approximate solution of the time-dependent Schr\"odinger equation \eref{eq-Schroedinger-equation} within the parameter subspace of the wave function \eref{eq-wave-function-parameter} is given by the McLachlan variational principle \cite{McLachlan1964}
\begin{equation}
	I = \| \ii \phi - \Hhat \psi \| \stackrel{!}{=} \mathrm{min.}
	\label{eq-McLachlan-variational-principle}
\end{equation}
where the quantity $I$ is minimized with respect to $\phi$ and $\phi=\dot{\psi}$ is set afterwards. For the parametrized wave function \eref{eq-wave-function-parameter} the application of the TDVP results in a set of first-order differential equations \cite{Fabcic2008}
\begin{equation}
	K \dot{\zz} = - \ii \vec{h},
	\label{eq-zdot}
\end{equation}
which determines the time evolution of the variational parameters, and where the matrix $K$ and the vector $\vec{h}$ are defined by ($m,n=1,\ldots,d$)
%
\begin{eqnarray}
	K_{mn} &= \int \! \dd^3\rr~ \left(\frac{\p \psi}{\p z_m}\right)^* \frac{\p \psi}{\p z_n} , \\
	{h}_m  &= \int \! \dd^3\rr~ \left(\frac{\p \psi}{\p z_m}\right)^* \hat{H}  \psi .
	\label{eq-def-K-h}%
\end{eqnarray}%
The energy of the system, which is given by the expectation value
\begin{equation}
	E(\zz) = \int \! \dd^3 \rr ~  \left( \psi(\rr,\zz(t)) \right)^* \, \hat{H} \, \psi(\rr,\zz(t))	,
	\label{eq-energy}
\end{equation}
is also a function of the variational parameters.

\subsubsection{Approach from field theory}
\label{sec-field-theory}

Within the McLachlan variational principle, equation \eref{eq-McLachlan-variational-principle}, one recognizes no obvious \AZ{Hamiltonian relation} between the energy functional \eref{eq-energy} and the equations of motion \eref{eq-zdot}. By contrast, such a relation becomes obvious, if we focus on the ansatz from field theory with the Hamiltonian density $\Hdensity=\psi^* \hat{H} \psi$. The energy functional is then given by
\begin{equation}
	E(\zz) = \int \! \dd^3 \rr ~ \Hdensity
	\label{eq-energy-field-theory}
\end{equation}
which is precisely the result of equation \eref{eq-energy}.

Within field theory, the dynamics of the wave function $\psi$ is given by (with the field momentum $\pi = \ii \psi^*$)
\begin{equation}
	\dot{\psi} = \frac{\p \Hdensity}{\ii \p \psi^*}
	\label{eq-equation-of-motion-psi-field-theory}
\end{equation}
which, for the variational ansatz \eref{eq-wave-function-parameter} of the wave function, leads to
\begin{equation}
	\left(\frac{\p \psi}{\p z_m}\right)^* \frac{\p\psi}{\p z_n} \dot{z}_n = - \ii \frac{\p \psi^*}{\p {z_m^*}} \hat{H} \psi
	\label{eq-zdot-without-int}
\end{equation}
and, after integration over $\dd^3 \rr$, is exactly the result of equations \eref{eq-zdot}--\eref{eq-def-K-h}. This step only needs a few lines of elementary calculations, including $\dot{\psi} = \frac{\p\psi}{\p z_n} \dot{z}_n$, $\p_{\psi^*} = \frac{\p z_m^*}{\p \psi^*} \p_{z_m^*}$, a multiplication with the inverse of the derivative $\left[ {\p z_m^*}/{\p \psi^*} \right] ^{-1} = {\p \psi^*}/{\p z_m^*}$ and the two assumptions that ${\p \psi^*}/{\p z_m^*} = \left({\p \psi}/{\p z_m}\right)^*$ and that the $z_m$ be complex, i.e.\ the derivatives with respect to $z_m$ and $z_m^*$ are independent. Moreover, we note that for $\psi^*$ the corresponding equation of motion $\dot{\psi}^* = -\p \Hdensity / \ii \p \psi$ yields the complex conjugate of equation \eref{eq-zdot-without-int}.

Vice versa, from equation \eref{eq-zdot-without-int} we see that the equations of motion derived from the TDVP do not only hold in the ``integrated version'',  equations \eref{eq-zdot}--\eref{eq-def-K-h}, but in these equations, we may set the integrands equal to one another, and those directly follow from the Hamilton operator $\hat{H}$ and its corresponding Hamiltonian density $\Hdensity$.

From this point of view, a ``Hamiltonian relation'' between the energy functional \eref{eq-energy} and the equations of motion \eref{eq-zdot} is obvious, and in addition, we obtain the conditions for the variational parameters to be \emph{complex} and the variational ansatz for the wave function to fulfil ${\p \psi^*}/{\p z_m^*} = \left({\p \psi}/{\p z_m}\right)^*$. The latter requirement directly follows from the Cauchy-Riemann differential equations, i.e.\ the variational ansatz for the wave function has to be complex differentiable in the variational parameters $z_m$ in order to obey this relation.

\subsection{Mapping to phase space}
\label{sec-mapping-to-phase-space}

Within the variational approach, the dynamics of the wave function and, with it, any ``reaction'' in the system, is given by the vector $\zz$ of the complex and time-dependent variational parameters. TST may then be applied to the wave packet if one is able to divide the space of the variational parameters -- after a suitable definition of the latter -- into ``reactants'' and ``products'', and if the point with the least energy on the dividing surface which forms the \AZ{activated complex} of the system is a stationary point of saddle-centre-\ldots-centre type of the energy functional \eref{eq-energy}. The corresponding reaction rate is then qualitatively given by the flux over this saddle in the complex $\zz$-space. However, in order to calculate the flux quantitatively a corresponding Hamiltonian in phase space is required.

The definition of the dividing surface and the calculation of the flux were given in reference \cite{Waalkens2008}, for the case that the corresponding Hamiltonian of the system is known. This is, however, not the case within the variational approach which is based on the energy functional \eref{eq-energy} and the equations of motion \eref{eq-zdot}. A globally defined Hamilton function which describes the system is \emph{not} known, in particular the energy functional \eref{eq-energy} cannot serve as a Hamiltonian because of the fact that the equations of motion \eref{eq-zdot} cannot be derived from the former via the canonical equations, and canonical variables are not even defined therein.

Nevertheless, as will be shown below it is possible to construct a \emph{local} Hamiltonian by an appropriate change of variables from the variational parameters $\zz$ to local phase space variables $\pp,\qq$ which equivalently describes the dynamics of the variational parameters and at the same time reproduces the energy of the system, given by equation \eref{eq-energy}. Therefore, this constructed Hamiltonian can be applied to calculating the flux.

In the following subsections, we provide the general concept of how to construct this equivalent Hamilton function. It consists of the subsequent steps:
\begin{enumerate}
	\item Taylor expand the equations of motion \eref{eq-zdot} in the vicinity of a fixed point $\zz_0$ in terms of the variational parameters and diagonalize the latter with respect to their linear part.

	\item Perform a normal form transformation of the Taylor expanded equations of motion and introduce canonical variables.

	\item Analogously to step (i), Taylor expand the energy functional \eref{eq-energy} and apply the change of coordinates which corresponds to the transformation from step (ii).

	\item Integrate the equations of motion according to Hamilton's equations and adapt the resulting Hamiltonian $\Htilde$ to the transformed energy functional to obtain the final Hamilton function $H$.
\end{enumerate}

\subsubsection{Transformation of the vector field}
\label{sec-trafo-vector-field}

We assume the differential equations \eref{eq-zdot} to exhibit a fixed point $\zz_0$. To obtain the local classical Hamilton function, we first Taylor expand these equations of motion in the vicinity of the fixed point up to the order $\nmax$ and split the expansion into its real and imaginary part. This results in the $2d$-dimensional real vector field
\begin{equation}
	\dot{\xx} = \vec{a}(\xx) = \sum_{n=1}^\nmax \vec{a}_n (\xx)
	\label{eq-DGL-series}
\end{equation}
where $\xx=(\mathrm{Re}(z_1-z_{01}), \mathrm{Im}(z_1-z_{01}), \ldots, \mathrm{Re}(z_d-z_{0d}), \mathrm{Im}(z_d-z_{0d}))$ denotes the deviation of the variational parameters from the fixed point and $\vec{a}_n (\xx)$ summarizes all terms of the expansion which are homogeneous of degree $n$. 

In order to \AZ{simplify} these differential equations we diagonalize them with respect to their linear part $\vec{a}_1(\xx) = A_1 \cdot \xx$ in the next step. Since this is trivial, we will not go into it in more detail. Note, however, that, even if it does not affect the result of the linearised and diagonalized equations of motion, this procedure itself is not unique, because one may work with arbitrary complex multiples of the eigenvectors of which one makes use of in this step. In contrast to the linearised differential equations, the choice of these multiples does affect higher order terms and the corresponding transformation of the energy functional which is explained in section \ref{sec-trafo-energy-adaptation}. We will use this freedom, among others, to adapt the ``integrated Hamiltonian'' to the transformed energy functional in a last step.

To further \AZ{simplify} the higher order terms of the remaining set of differential equations, we perform a near-identity transformation $\xx\to\yy$ given by (cf.\ reference [13])
\begin{equation}
	\xx = \vec{\phi}_\eps(\yy), \qquad \xx = \vec{\phi}_{\eps=0}(\yy) = \yy
	\label{eq-change-of-coordinates}
\end{equation}
where the parameter $\eps$ is chosen in such a way that we obtain the identity-transformation for $\eps=0$ and $\xx = \vec{\phi}_\eps(\yy)$ solves the differential equation 
\begin{equation}
	\frac{\dd \xx}{\dd \eps} = \vec{g} (\xx)
	\label{eq-generator}
\end{equation}
with $\generator(\xx)$ being the generating function. Concerning the normal form transformations which will be applied in the following, we keep close to the notation introduced in reference \cite{Murdock2010} and also use the notion ``normal form'' as defined therein.

Under the change of variables \eref{eq-change-of-coordinates}, given by a specific generating function $\generator(\xx)$ the vector field \eref{eq-DGL-series} transforms as \cite{Murdock2010}
\begin{equation}
	\dot{\yy} = \vec{b}(\yy) = \sum_{k=0}^\infty \strich{ \frac{1}{k!} \Lop{\generator}^k \vec{a} (\xx) }_{\xx=\yy}
	\label{eq-trafo-of-vector-field}
\end{equation}
with the Lie-operator $\Lop{\generator}$ defined by
\begin{equation}
	\Lop{\generator}\vec{a}(\xx) = \Bigl[ \generator(\xx) \cdot \nabla \Bigr] \vec{a}(\xx) - \Bigl[ \vec{a}(\xx) \cdot \nabla \Bigr] \generator(\xx).
	\label{eq-Lie-definition}
\end{equation}

By simply expanding the sum in equation \eref{eq-trafo-of-vector-field} and sorting by the polynomial order $n$ it can, with $\vec{b}(\yy)=\sum_n \vec{b}_n(\yy)$, easily be shown that a generating function $\generator_n (\xx)$ which is homogeneous of degree $n$ transforms the corresponding term $\vec{a}_n(\yy)$ according to the homological equation
\begin{equation}
	\vec{b}_n (\yy) = \vec{a}_n (\yy) + \Lop{\generator_n} \vec{a}_1 (\yy).
	\label{eq-trafo-bn}
\end{equation}
Terms of $\vec{a}(\yy)$ of lower degree than $n$ remain unchanged. Therefore, by choosing a specific degree $n$ of the generator $\generator_n$, we can normal transform the vector field $\vec{a}(\yy)$ order by order.

Introducing the expansions
\begin{eqnarray}
	\vec{a}_n(\yy) &&= \sum_{\abs{\mm}=n} \vec{\alpha}_{\mm} \yy^{\mm}, 	\\	
	\vec{b}_n(\yy) &&= \sum_{\abs{\mm}=n} \vec{\beta }_{\mm} \yy^{\mm}, 	\\
	\vec{g}_n(\yy) &&= \sum_{\abs{\mm}=n} \vec{\gamma}_{\mm} \yy^{\mm},
\end{eqnarray}
where we made use of the multi-index notations $\xx^{\mm}=x_1^{m_1} x_2^{m_2}x_3^{m_3} \ldots $ and $\abs{\mm}=\sum_i m_i$ with the integer vector $\mm$, and inserting these into equation \eref{eq-trafo-bn} the transformation can be directly applied to the single coefficients which componentwise transform according to
\begin{equation}
	\beta_{\vec{m}i} = \alpha_{\vec{m}i} + ( \vec{\lambda} \vec{m} - \lambda_i ) \gamma_{\vec{m}i} \, .
	\label{eq-trafo-coefficients}
\end{equation}
Here, the $\lambda_i$ are the eigenvalues of the linearised equations of motion. In case of $\vec{\lambda} \vec{m} - \lambda_i \neq 0$ the respective monomial $\alpha_{\vec{m}i}$ can be eliminated ($\beta_{\vec{m}i}=0$) by an appropriate choice of $\gamma_{\vec{m}i}$ whereas in the \AZ{resonant} case 
\begin{equation}
	\vec{\lambda} \vec{m} - \lambda_i = 0
	\label{eq-resonance-condition}
\end{equation}
it remains unchanged ($\beta_{\vec{m}i} = \alpha_{\vec{m}i}$) and \emph{cannot} be eliminated. Hence, the eigenvalues of the linearised equations of motion determine the whole structure of the normal form expansion. 

Since the equations of motion derived from the TDVP, equations \eref{eq-zdot}--\eref{eq-def-K-h}, are invariant under time reversal $t \to -t$ (together with $(\psi,\zz)\to(\psi^*, \zz^*)$), the eigenvalues of the linearised equations of motion in general exhibit the structure
\begin{equation}
	\lambda_i = \pm \delta_1, \, \ldots,\, \pm \delta_j,\, \pm \ii \omega_{j+1}, \, \ldots  , \, \pm \ii \omega_d 
	\label{eq-eigenvalue-structure}
\end{equation}
with $\delta_i, \omega_i \in \mathbbm{R}$, i.e.\ they occur pairwise with different sign. Assuming rational independence of the eigenvalues  $\delta_i, \ii\omega_i~(i=1,2,\ldots,d)$ this structure directly implies that equation \eref{eq-resonance-condition} is fulfilled for 
\begin{eqnarray}
	m_{2i-1} &= m_{2i} + 1, \qquad	& (i=1,2,\ldots,d), \label{eq-condition-m-a}\\
	m_{2j-1} &= m_{2j},				& (j\neq i).		\label{eq-condition-m-b}%
\end{eqnarray}%
Consequently, the form of the normal form transformed equations of motion is fixed because only monomials satisfying equations \eref{eq-condition-m-a} and \eref{eq-condition-m-b} will remain. We now assume these equations of motion to be sorted in a way that the linear terms occur blockwise according to the eigenvalue structure \eref{eq-eigenvalue-structure} where each block $\dot{x}_{2i-1},\dot{x}_{2i}$ ($i=1,\ldots,d$) contains the same eigenvalue differing only in its sign. Then they take the form
\begin{eqnarray}
	\dot{x}_{2i-1} &= 	\sum_{\mm}	\beta_{\mm (2i-1)}	x_{2i-1}^{m_{2i-1}} x_{2i}^{m_{2i}-1} 	\prod_{j\neq i} (x_{2j-1} x_{2j})^{m_{2j}},	
	\label{eq-structure-monomials-x-a}\\
	\dot{x}_{2i} &=  	\sum_{\mm} 	\beta_{\mm (2i)~~~} x_{2i-1}^{m_{2i-1}-1} x_{2i}^{m_{2i}} 	\prod_{j\neq i} (x_{2j-1} x_{2j})^{m_{2j}}.
	\label{eq-structure-monomials-x-b}
\end{eqnarray}

We now systematically introduce ``canonical'' coordinates $ \tilde{q}_i = x_{2i-1}$ and momenta $\tilde{p}_i = x_{2i}$ ($i=1,\ldots,d$) with the help of which we can rewrite equations \eref{eq-structure-monomials-x-a} and \eref{eq-structure-monomials-x-b} as
\begin{eqnarray}
	\dot{\tilde{q}}_i&=   	\sum_{\mm} 	\beta_{\mm (2i-1)}  \tilde{q}_i^{m_{2i-1}} \tilde{p}_i^{m_{2i}-1} \prod_{j\neq i} (\tilde{q}_j \tilde{p}_j)^{m_{2j}},	
	\label{eq-structure-monomials-qp-a}\\
	\dot{\tilde{p}}_i&=  	\sum_{\mm} 	\beta_{\mm (2i)}~~~ \tilde{q}_i^{m_{2i-1}-1} \tilde{p}_i^{m_{2i}} \prod_{j\neq i} (\tilde{q}_j \tilde{p}_j)^{m_{2j}}.
	\label{eq-structure-monomials-qp-b}%
\end{eqnarray}%

\subsubsection{Transformation of the energy functional}
\label{sec-trafo-energy-adaptation}

To obtain the local energy functional in phase space coordinates $\tilde{\qq},\tilde{\pp}$, we need to apply the change of coordinates corresponding to the normal form transformation of the equations of motion also to the expanded energy functional
\begin{equation}
	E = \sum_{n=0}^{\nmax+1} E_n (\xx).
	\label{eq-energy-expansion}
\end{equation}

Since the generating function $\generator_n$ is, for each order, known from the normal form transformation of the vector field $\vec{a}(\xx)$, the explicit change of coordinates corresponding to the transformation \eref{eq-trafo-bn} can easily be obtained. In dependence of the generating function it reads \cite{Murdock2010}
\begin{equation}
	\yy = \sum_{k=0}^\infty \frac{1}{k!} \Dop{\generator}^k \xx,
\end{equation}
where the ``right-hand multiplication operator'' $\Dop{\generator}$ is defined by its action
\begin{equation}
	\Dop{\generator} \ff(\xx) = \frac{\dd\ff(\xx)}{\dd \xx} \cdot \generator (\xx)
	\label{eq-Doperator-definition}
\end{equation}
on differentiable functions $\ff(\xx)$.

Inserting the transformation corresponding to each step into the expanded energy functional \eref{eq-energy-expansion} yields the form
\begin{equation}
	E = \sum_{\mm}^\infty \xi_{\mm} \prod_{j} (\tilde{q}_j \tilde{p}_j)^{m_j}.
	\label{eq-energy-series-J}
\end{equation}
The fact that the energy functional can be expressed in products $\tilde{q}_j \tilde{p}_j$, is here due to the ``Hamilton-like'' relation between the equations of motion derived from the TDVP and the energy functional, explained in section \ref{sec-field-theory}. 

\subsubsection{Construction of the Hamilton function $H$}
\label{sec-adaptation}

The exponents of the $\tilde{q}_i, \tilde{p}_i$ in the $i$-th component of the transformed equations of motion  \eref{eq-structure-monomials-qp-a}--\eref{eq-structure-monomials-qp-b} exactly differ by one and do not differ for $\tilde{q}_j, \tilde{p}_j$ ($j\neq i$) which is, both, due to the eigenvalue structure \eref{eq-eigenvalue-structure} and leads to the fact that they can easily be integrated to a common Hamilton function
\begin{equation}
	\Htilde = \sum_{i,\mm} \frac{\beta_{\mm (2i)}}{m_{2i}} (\tilde{q}_i \tilde{p}_i)^{m_{2i}} \prod_{j\neq i} (\tilde{q}_j \tilde{p}_j)^{m_{2j}}
	\label{eq-Htilde}
\end{equation}
according to Hamilton's equations
\begin{equation}
	\dot{\tilde{q}}_i = \frac{\p \Htilde}{\p \tilde{p}_i}, \qquad \dot{\tilde{p}}_i = - \frac{\p \Htilde}{\p \tilde{q}_i}.
\end{equation}
This is possible, if the remaining coefficients $\beta_{\mm i}$, after the normal form expansion, satisfy the conditions of integrability
\begin{eqnarray}
	\beta_{\mm (2i-1)} &= - \beta_{\mm (2i)}, 		\label{eq-integration-conditions-a}				\\
	\frac{\beta_{\mm (2i-1)}}{m_{2i}} &= \frac{\beta_{\mm (2i'-1)}}{m_{2i'}},\label{eq-integration-conditions-b}	\\
	\frac{\beta_{\mm (2i)}}{m_{2i-1}} &= \frac{\beta_{\mm (2i')}}{m_{2i'-1}}
	\label{eq-integration-conditions-c}
\end{eqnarray}
for all $i,i'=1,\ldots,d$ ($i\neq i'$). Note that equation \eref{eq-integration-conditions-a} is automatically fulfilled if the diagonalisation of the equations of motion is performed with, in each case, two corresponding complex conjugate eigenvectors which are normalized identically. However, the equations \eref{eq-integration-conditions-b}--\eref{eq-integration-conditions-c} are, in general, not automatically satisfied after the transformations performed above.

In a system with one degree of freedom ($d=1$) only the condition \eref{eq-integration-conditions-a} plays a role, and the formal integration to the Hamiltonian $\Htilde$ can immediately be carried out. The tilde on the integrated Hamilton function \eref{eq-Htilde}, however, is intended to emphasize that, even if the equations of motion after the normal form expansion can be derived from it according to Hamilton's equations, it will, in general, not represent the energy of the system.

In order to achieve both, i.e.\ the satisfaction of the conditions of integrability \eref{eq-integration-conditions-b}--\eref{eq-integration-conditions-c} for systems with $d\geq2$ degrees of freedom and the local equivalence of the integrated Hamiltonian with the energy functional, a final transformation $\tilde{\qq} \to \qq$ and $\tilde{\pp} \to \pp$ is necessary. The latter makes use of the fact that the transformations performed in sections \ref{sec-trafo-vector-field} and \ref{sec-trafo-energy-adaptation} still leave some freedom, e.g.\ the choice of multiples of eigenvectors when diagonalizing the equations of motion.

For this purpose, we scale the generalized coordinates $\tilde{\qq} \to \qq$ and the momenta $\tilde{\pp} \to \pp$ in  \eref{eq-structure-monomials-qp-a}--\eref{eq-structure-monomials-qp-b} and \eref{eq-energy-series-J} with some time-\emph{in}dependent functions $\nu_{q_i}(\qq,\pp)$ and $\nu_{p_i}(\qq,\pp)$ according to
\begin{equation}
	\tilde{q}_i = \nu_{q_i}(\qq,\pp)\, q_i~,	\qquad
	\tilde{p}_i = \nu_{p_i}(\qq,\pp)\, p_i.
	\label{eq-nu-scaling}
\end{equation}
Since the $\tilde{q}_i$ and $\tilde{p}_i$ do only occur as products $\tilde{p}_i \tilde{q}_i$ in the integrated Hamiltonian $\Htilde$ as well as in the transformed energy functional \eref{eq-energy-series-J}, the precise form of the $\nu_{q_i}, \nu_{p_i}$ is not of interest. We only demand their product
\begin{equation}
	\mu_i (\qq,\pp) = \nu_{q_i}(\qq,\pp) \, \nu_{p_i}(\qq,\pp)  = 1 + \sum_{\mm} \mu_{\mm} \prod_j (q_j p_j)^{m_j}
	\label{eq-mu-power-series}
\end{equation}
also to be a power series of the products $q_j p_j$ which is necessary in order to preserve the structure of the respective equations.

Because both sides of the equations \eref{eq-structure-monomials-qp-a} and \eref{eq-structure-monomials-qp-b} are scaled according to equation \eref{eq-nu-scaling}, one of the functions $\nu_{q_i}(\qq,\pp)$ and $\nu_{p_i}(\qq,\pp)$, respectively, cancels out, and the remaining products $\nu_{q_i}\nu_{p_i}$ can be replaced by the expansion \eref{eq-mu-power-series}. After rearranging the right-hand sides of the respective equations the coefficients of the expansions then depend on the $\mu_{\mm}$ and the conditions of integrability \eref{eq-integration-conditions-b}--\eref{eq-integration-conditions-c} provide a linear system of equations for their determination. The latter, however, is underdetermined and to obtain a complete set of equations one also requires the equivalence of the integrated Hamiltonian and the transformed energy functional.

The cancelling out of one $\nu_{q_i}(\qq,\pp)$ or $\nu_{p_i}(\qq,\pp)$ in the equations of motion leads to the crucial result that the occurrence of the $\mu_{\mm}$ in the integrated Hamiltonian $\Htilde$ and in the energy functional exactly differs by one order,
\begin{eqnarray}
\fl	\Htilde =\,	&  \Htilde_0 + \Htilde_1(\qq,\pp) + \Htilde_2(\qq,\pp,\mu_{\mm, \abs{\mm}=2})  +\Htilde_3(\qq,\pp,\mu_{\mm, \abs{\mm}\leq 3}) + \ldots\,,	\label{eq-mu-occurance-H}\\[.5em]
\fl	E 	=\,	&  E_0 + E_1(\qq,\pp,\mu_{\mm, \abs{\mm}=2}) + E_2(\qq,\pp,\mu_{\mm, \abs{\mm} \leq 3}) 
			  +E_3(\qq,\pp,\mu_{\mm, \abs{\mm} \leq 4}) + \ldots\,,
	\label{eq-mu-occurance-E}%
\end{eqnarray}
i.e.\ the order $n$ term $E_n$ of the transformed energy functional contains those $\mu_{\mm}$ with $\abs{\mm}\leq n+1$, whereas in the corresponding term of the integrated Hamiltonian $\Htilde$ only the scaling coefficients $\mu_{\mm}$ with $\abs{\mm}\leq n$ occur. 

Consequently, orderwise comparing the coefficients of the expansions \eref{eq-mu-occurance-H}--\eref{eq-mu-occurance-E}, yields both the constant of integration $\Htilde_0$ and the required equations to uniquely determine the $\mu_{\mm}$. The latter are, finally, determined orderwise as the solutions of a linear system of equations considering that the conditions of integrability \eref{eq-integration-conditions-b}--\eref{eq-integration-conditions-c} and the requirement of the equations \eref{eq-mu-occurance-H}--\eref{eq-mu-occurance-E} to be identical are fulfilled simultaneously.

After this proper choice of the $\mu_{\mm}$ and the definition of action-variables by $J_i={q}_i {p}_i$ if ${J}_i$ corresponds to a real eigenvalue $\lambda_i$ and ${J}_i= \ii {q}_i {p}_i$ in case of purely imaginary ones $\ii \omega_i$, we have
\begin{equation}
	H \left(\vec{J} \right) = E\left(\vec{J}\right) = \Htilde \left(\vec{J}\right).
	\label{eq-Hamiltonian}
\end{equation}
This serves as classical Hamilton function in the sense that it reproduces the energy of the system, given by equation \eref{eq-energy}, and at the same time its Hamiltonian equations of motion equivalently to equation \eref{eq-zdot} describe the dynamics in the vicinity of the fixed point up to some previously chosen order.

\subsection{Thermal decay rates}

After the classical Hamilton function \eref{eq-Hamiltonian} has been constructed, classical TST can be applied. In this section, we review the formulas to calculate the thermal decay rates which follow from this theory.

We now assume the Hamilton function in phase space to exhibit a local minimum corresponding to its metastable ground state. Moreover let there exist an equilibrium point of saddle-centre-\ldots-centre type which is the only channel the system may decay.

The decay rate at a fixed energy is then given by the flux through a dividing surface which divides the phase space into a region of \AZ{reactants} and \AZ{products}, respectively (see reference \cite{Waalkens2008}). If the system is in contact to a bath of finite temperature, the thermal decay rate is the Boltzmann average of this flux.

\subsubsection{The classical case}

The directional flux through the dividing surface between the reactant's and product's region in phase space for a fixed energy is \cite{Waalkens2008, Waalkens2004, MacKay1990}
\begin{equation}
	f(E) = (2\pi)^{d-1} \mathcal{V}(E),
	\label{eq-flux-over-saddle}
\end{equation}
with $\mathcal{V}(E)$ being the phase space volume of the actions $(J_2,\ldots,J_d)$ which is enclosed by the contour $H_\mathrm{CNF}(0,J_2, \ldots, J_d) \leq E$ and $J_1=0$ corresponding to the \AZ{unstable direction} of the saddle. The thermal decay rate is then given by the Boltzmann average of equation \eref{eq-flux-over-saddle} which, after a short calculation, yields
\begin{equation}
	\Gamma_\mathrm{cl} = \frac{1}{2\pi \hbar ^d \beta Z_0} \int  \ee^{-\beta H_\mathrm{CNF}(0,J_2,\ldots,J_d)} \dd J_2 \ldots \dd J_d,
	\label{eq-decay-rate-cl}
\end{equation}
where $\beta=1/k_\mathrm{B}T$, and $Z_0$ is the canonical partition function (cf.\ reference \cite{Toller1985}). Since nearly all states will be localized in the vicinity of the local minimum, we will approximate the latter by 
\begin{equation}
	Z_0 = \frac{1}{\hbar^d} \int \dd J_1\, \ldots \dd J_d ~ e^{-\beta H'_ \mathrm{CNF}(J'_1,\ldots,J'_d)}
	\label{eq-partition-function-cl}
\end{equation}
with $H'_ \mathrm{CNF}(J'_1,\ldots,J'_d)$ being the normal form expansion at the local minimum.

In general, both integrals in equations \eref{eq-decay-rate-cl} and \eref{eq-partition-function-cl} will not converge, which is due to the fact that, when actually computing the normal form, the expansion has to be truncated at some order. To compensate this, we restrict the phase space volume over which is integrated to the condition 
\begin{equation}
	\omega_i = \frac{\p H_\mathrm{CNF}}{\p J_i} \geq 0, \qquad
	i=1,\ldots,d
	\label{eq-condition-frequencies}
\end{equation}
taking into account the fact that all frequencies occurring on the tori in phase space have to be non-negative.

\subsubsection{The quantum case}

For the quantum mechanical case in equation \eref{eq-normal-form-6}, we proceed analogously. The quantum mechanical reaction probability for a one-degree-of-freedom system at a fixed energy is given by \cite{Schubert2006, Waalkens2008}
\begin{equation}
	N(E) = \frac{1}{1+\ee^{-2\pi J(E)/\hbar}}
	\label{eq-qm-reaction-probability}
\end{equation}
where the action variable $J$ is implicitly given by $H_\mathrm{QNF}(J) = E$.

Again the thermal decay rate $\Gamma_\mathrm{qm}$ is given by the Boltzmann average of equation \eref{eq-qm-reaction-probability} which, after integration by parts, reads
\begin{equation}
	\Gamma_\mathrm{qm} = \frac{1}{\hbar^2\beta Z_0} \int \dd J ~ \frac{\ee^{-\beta H_\mathrm{QNF}(J)-2\pi J(E) /\hbar}}{(1+\ee^{-2\pi J(E)/\hbar})^2}.
	\label{eq-decay-rate-qm}
\end{equation}

In the quantum case the canonical partition function is evaluated as a sum
\begin{equation}
	Z_0 = \sum_n \ee^{-\beta H'_\mathrm{QNF}(J'=\hbar(n+1/2))},
	\label{eq-partition-function-qm}
\end{equation}
and because of the same reasons as in the classical case, we restrict the phase space volume for the integration in equation \eref{eq-decay-rate-qm} and the summation in equation \eref{eq-partition-function-qm} to the condition \eref{eq-condition-frequencies} with $H_\mathrm{CNF} \to H_\mathrm{QNF}$.

\section{Application and results}
\label{sec-results}

\subsection{The cubic model potential}

\begin{figure}[t]
	\centering
	\includegraphics[width=.5\columnwidth]{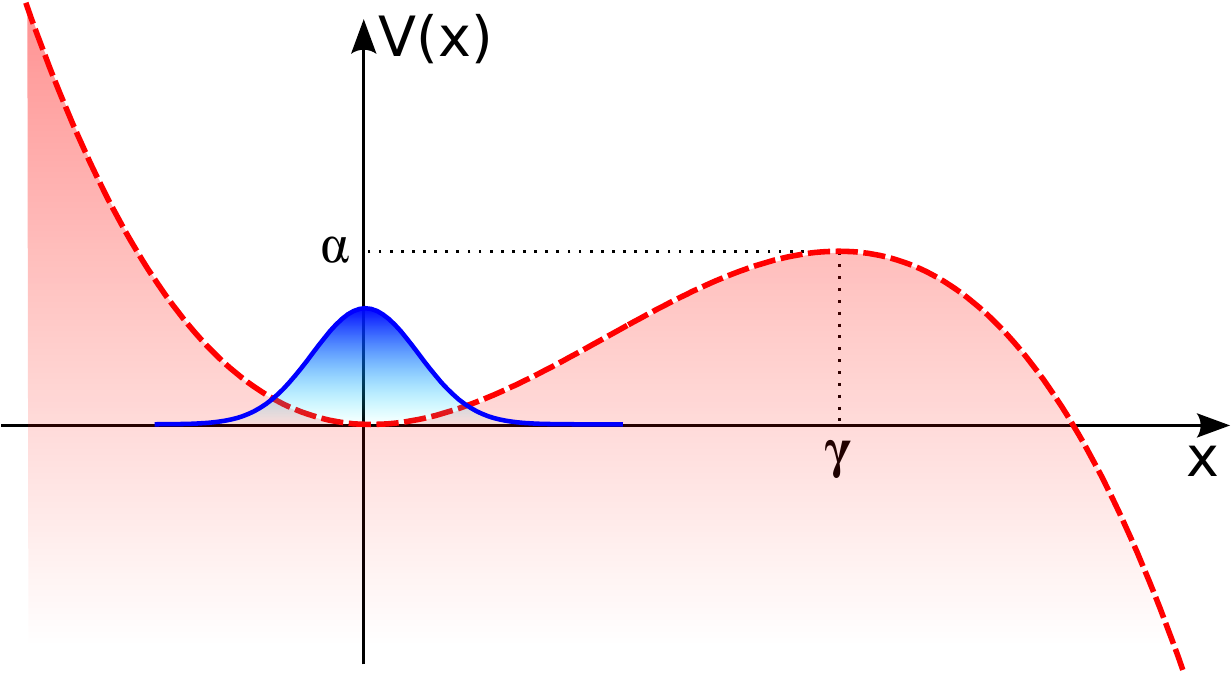}
	\caption{Schematic drawing of the cubic model potential for $\alpha,\gamma>0$ (dashed line). The local minimum at Min$(0,0)$ is the metastable ground state of the system. A particle in the ground state can, after sufficient thermal excitation, cross the barrier formed by the local maximum at Max$(\gamma, \alpha)$ and will then escape to $x\to \infty$. Within the variational approach, the particle is described by a Gaussian wave function of fixed width (solid line), and its two stationary states are located at the minimum and the saddle.}
	\label{fig-cubic-potential}
\end{figure}

To demonstrate the applicability of the procedure to calculating thermal decay rates within the framework of a variational approach to the quantum mechanical wave function, we consider a cubic model potential 
\begin{equation}
	V(x) = - \frac{\alpha}{\gamma^3} x^2 (2x-3\gamma).
	\label{eq-cubic-potential}
\end{equation}
It features a local minimum at Min$(0,0)$ and a local maximum at Max$(\gamma, \alpha)$ for $\alpha, \gamma >0$ (see figure \ref{fig-cubic-potential}).

Within the variational approach, we approximate the particle's wave function by a frozen Gaussian
\begin{eqnarray}
	\psi(x,t) &= \tilde{\mathcal{N}} 	\exp \left( -a(x-q)^2+\ii p(x-q) \right) \\	
			&= \mathcal{N} 		\exp \left( -ax^2+zx \right) \label{eq-frozen-gaussian-b}
	\label{eq-frozen-gaussian}%
\end{eqnarray}%
with fixed width $a>0$. The centre $q$ and momentum $p$ of the wave function are combined in the complex variational parameter $z$ with real and imaginary part $z_r=2aq$ and $z_i=p$, respectively, and $\mathcal{N}=\left( \pi / 2a \right)^{1/4} \exp\left(z_r^2/4a \right) = \tilde{\mathcal{N}} \exp \left(-aq^2-\ii pq \right)$ gives the normalization of the wave function: $\int\! \dd x \, \abs{\psi(x,t)}^2 =1$. With this definition every point in the complex $z$-plane can be identified with a certain position and state of motion of the Gaussian wave packet and vice versa.

In this paper, we use a single frozen Gaussian, since in this case the energy functional \eref{eq-energy} and the equations of motion for the variational parameters \eref{eq-zdot} can be obtained explicitly. Note, however, that the procedure is neither limited to \emph{Gaussian} wave functions nor to the use of a \emph{single} Gaussian, but can without any change also be applied to wave packets of \emph{coupled} Gaussians (cf. reference \cite{Junginger2011c}), or of different form.

The energy functional of the system is given by the expectation value of the Hamiltonian ($\hbar=m=1$)
\begin{equation}
	H = - \frac{1}{2} \Delta + V(x)
\end{equation}
with the potential $V(x)$ in equation \eref{eq-cubic-potential}. For the wave function \eref{eq-frozen-gaussian-b} it reads
\begin{equation}
	E = \frac{\alpha  \left(3 a^2 \gamma +3 a z_r (\gamma z_r - 1)- z_r^3 \right)}{4 a^3 \gamma ^3}+\frac{1}{2} \left(a + z_i^2\right).
	\label{eq-energy-functional}
\end{equation}
The application of the TDVP gives the equations of motion for the time evolution of the variational parameter and, after splitting into real and imaginary part, yields
\begin{eqnarray}
	\dot{z}_r 	=& 	2 a z_i,	
	\label{eq-zdot-model-potential-a}\\ 
	\dot{z}_i 	=&	\frac{3 \alpha  z_r^2}{2 a^2 \gamma ^3}-\frac{3 \alpha  z_r}{a \gamma ^2}+\frac{3 \alpha }{2 a \gamma ^3}.
	\label{eq-zdot-model-potential-b}
\end{eqnarray}

TST can be applied to calculating the reaction rate of this system because of the following reason: First, we define ``reactants'' as initial states of motion for which the wave packet remains in the potential well, i.e.\ $x<\gamma$ for $t\to \infty$, and ``products'' as those which escape to $x\to\infty$ for $t\to \infty$. With this definition the complex $z$-plane can be divided, by solving the equations of motion \eref{eq-zdot-model-potential-a}--\eref{eq-zdot-model-potential-b}, in a region of reactants and products, respectively. The reaction rate is then given by the flux through the dividing surface between these two regions.

Thermal excitation now enables a wave packet corresponding to the reactants to cross the barrier of the potential $V(x)$ and to become a product. For low temperatures, the reaction path will cross the dividing surface at its point with the least energy. This point in the $z$-plane is a saddle of the energy functional \eref{eq-energy-functional}, consequently the decay rate of the system is given by the flux over this saddle.

However, since $z$ is a ``bad'' variable in the sense that -- as already stated in section \ref{sec-mapping-to-phase-space} -- the equations of motion  \eref{eq-zdot-model-potential-a}--\eref{eq-zdot-model-potential-b} cannot be derived from the energy functional by the canonical equations, the latter cannot be regarded as a Hamilton function which allows for the calculation of the flux.

In order to obtain a Hamilton function $H(J)$ which can be used for this purpose, we perform the steps described in section \ref{sec-mapping-to-phase-space}, i.e.\ we determine the fixed points of the equations \eref{eq-zdot-model-potential-a}--\eref{eq-zdot-model-potential-b}, expand them in their vicinity and order by order perform a normal form transformation. After integrating and adapting the result to the transformed energy functional, we end up with the desired Hamiltonian \eref{eq-Hamiltonian} which is locally equivalent to the equations \eref{eq-energy-functional}--\eref{eq-zdot-model-potential-b}. To illustrate the procedure in a more descriptive way, we provide a numerical example in the appendix for a certain set of parameters.

Note that -- as it is also the case in the analogous classical system of a point particle in the potential $V(x)$ -- the equations of motion \eref{eq-zdot-model-potential-a}--\eref{eq-zdot-model-potential-b} exhibit two fixed points. One of them corresponds to the metastable ground state of the system and the other to an unstable excited state, whose corresponding wave function is located in the vicinity of the potential's local maximum. While the ground state also exists in the framework of the exact Schr\"odinger equation, the existence of the unstable excited state is a consequence of the ansatz of the wave function within the variational approach. A wave packet which is located at the maximum of the potential $V(x)$ is, of course, not a stationary solution of the exact Schr\"odinger equation, but would dissolve. Within the variational approach, this dissolving character, however, is taken into account by the instability of this state.

Note furthermore, that the eigenvalues of the linearised equations of motion at the fixed points, which crucially determine the normal form expansion and, with it, the constructed Hamiltonian $H(J)$, approximately reproduce the eigenfrequency of the harmonic oscillator which one obtains by harmonically approximating the potential $V(x)$ at its minimum and maximum, respectively. For wave functions with smaller extension (larger $a$) the eigenvalues converge to the oscillation frequency of the harmonic oscillator and in the limit $a\to \infty$ exactly reproduce it as is expected, since the dynamics of a strongly localized wave packet will be the same as that of a point particle.

\subsection{Comparison of the decay rates}

\begin{figure}[t]
	\centering
	\includegraphics[width=0.7\columnwidth]{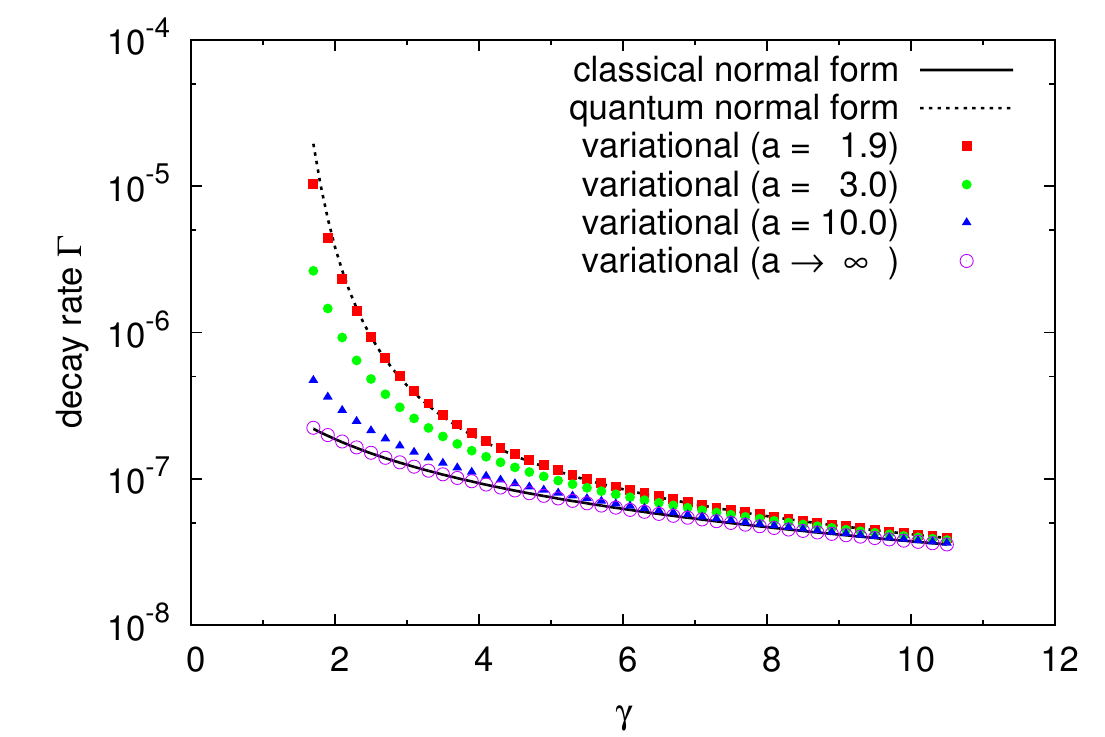}
	\caption{Comparison of the thermal decay rates of a particle placed in the metastable ground state of the cubic potential calculated by the classical normal form (solid line), the quantum normal form (dashed line) and the variational approach with a frozen Gaussian wave function of different width $a$ (dots). The data have been calculated in dependence of the parameter $\gamma$, for a temperature of $\beta = 1/k_\mathrm{B} T= 1.5$ and a barrier height of $\alpha=10$ in equation \eref{eq-cubic-potential}. In the limit $a\to \infty$ of a very narrow wave function (empty circles) the  variational ansatz covers the results of the classical normal form while the decay rates obtained from this method increase with decreasing width parameter $a$ (triangles and filled circles). For a broad wave function (squares) it shows a perfect match with the decay rate obtained from the quantum normal form of the corresponding point particle.}
	\label{fig-different-width-parameter}
\end{figure}

In order to compare the thermal decay rates calculated by the different methods we first determine the quantum and the classical normal form of a point particle in the model potential $V(x)$ from equation \eref{eq-normal-form-6} and its normal form analogue of the variational parameter space as described in section \ref{sec-mapping-to-phase-space}. The thermal decay rates are then given by equation \eref{eq-decay-rate-cl} for the classical case and equation \eref{eq-decay-rate-qm} for the quantum case (for the latter we set $\hbar=1$ throughout).

Figure \ref{fig-different-width-parameter} shows a comparison of the thermal decay rates calculated for the cubic potential by the three methods in dependence of the parameter $\gamma$ of the potential and for different width parameters $a$ of the Gaussian trial wave function. The temperature is set to $\beta=1.5$ and we use a barrier height of $\alpha=10$ in equation \eref{eq-cubic-potential}. 

The classical (solid line) and quantum (dashed line) normal forms reveal a significantly differing decay rate which is especially pronounced for small $\gamma$ due to the quantization effects in the narrow potential well. The variational approach with a single Gaussian trial wave function, however, is able to reproduce both curves. In the limit of a very narrow wave function (empty circles), i.e.\ $a\to \infty$, we find a perfect match of the latter with the classical result. When we increase the width of the Gaussian wave function, also the decay rate increases (triangles and filled circles). For a width of $a=1.9$ (squares), we finally recover the decay rate calculated from the quantum normal form of the corresponding point particle.

The classical limit of the decay rate can easily be understood considering the classical limit of the variational wave function. For the narrow wave function, only the local properties of the potential will be of importance so that the potential can be approximated harmonically in this limit. The characteristic width of the Gaussian is then obtained from that of the ground state wave function of the harmonic oscillator, which is given by $a=\omega/(2\hbar)$, with $\omega=\sqrt{|V''(x)|}_{x=0}$. The classical limit $\hbar\to0$ then yields $a\to \infty$ for which we observe convergence of the decay rate obtained from the variational approach to the one obtained from the classical normal form. For broad wave functions the anharmonicity of the potential becomes relevant and the width of the harmonic oscillator will be only a crude approximation. For $\hbar=1$ the width of $a=1.9$, for which we observed good agreement with the quantum normal form result, is in accordance with the width of the harmonic oscillator ground state within about a factor of $2$ over the whole range. According to this, intermediate values $1.9<a<\infty$ correspond to effective values $0<\hbar<1$ of the Planck constant.

\begin{figure}[t]
	\centering
	\includegraphics[width=0.7\columnwidth]{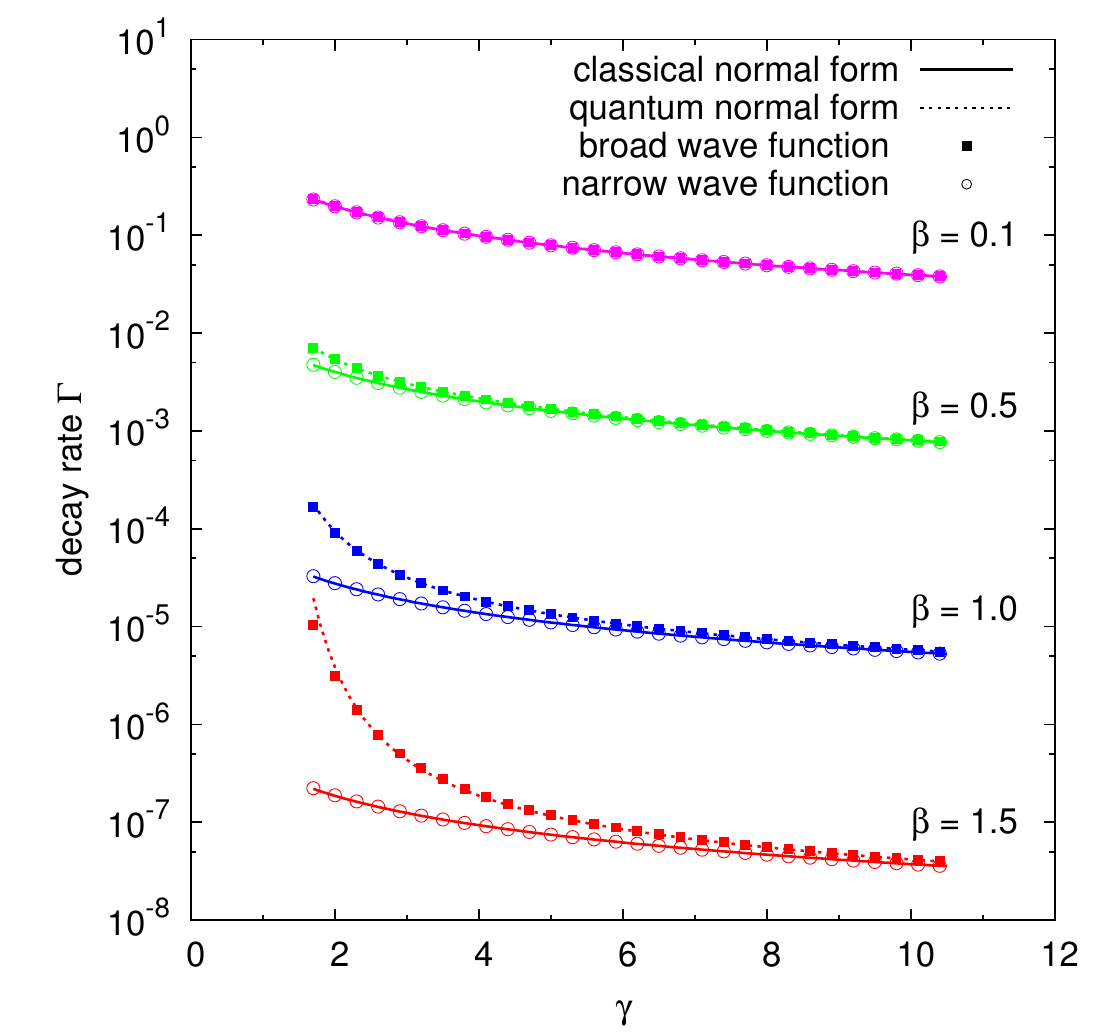}
	\caption{Thermal decay rates for the cubic potential for different temperatures. As in figure \ref{fig-different-width-parameter} we compare the results of the classical (solid lines) and quantum (dashed lines) normal forms as well as the variational approach in the limit of broad (squares) and narrow (empty circles) Gaussian trial wave functions. With increasing temperature, also the decay rates increase and the difference between the classical and quantum normal forms becomes smaller in the same way as the two cases of narrow and broad wave function do within the variational approach. For $\beta \lesssim 0.1$ the decay rates calculated by the different methods cannot be distinguished any more.}
	\label{fig-different-temp}
\end{figure}

We further observe this behaviour at different temperatures as can be seen in figure \ref{fig-different-temp} for some exemplary values. Again, the results using the classical normal form (solid lines) are perfectly recovered within the limit $a\to\infty$ of very narrow wave functions (empty circles) and for broad ones (squares) the decay rates agree with those obtained from quantum normal form (dashed lines). 

With increasing temperature, the calculated decay rates also increase rapidly by several orders of magnitude. However, the difference between the classical and the quantum normal forms becomes smaller as well as it does for the two limits of the wave function concerning their width. At high temperatures compared to the height of the potential barrier we reach the classical limit. For the parameters used here, this is the case for $\beta \lesssim 0.1$, where the difference of the decay rates calculated by the three methods vanishes and they cannot be distinguished any more.

We emphasize that -- in contrast to the classical and the quantum normal form -- the method with wave packets described by the TDVP does \emph{not} require a Hamilton function for the system to be known at the beginning of the procedure. Instead, only an expression for the energy functional and a set of differential equations describing the time evolution of the variational parameters are necessary. The required Hamiltonian is, after the mapping to phase space coordinates, a natural result of the transformations discussed in this paper.

\section{Summary and outlook}

We have presented a method to apply transition state theory to wave packet dynamics in metastable Schr\"odinger systems, which is based on local expansions of the energy functional and the equations of motions derived from a time-dependent variational principle in the vicinity of the barrier fixed point of saddle-centre-\ldots-centre type. Fulfilling the integrability conditions, the equations of motion can, after a normal form expansion, be integrated to a common Hamilton function expressed in action-variables. Its equivalence to the energy functional is guaranteed by an additional transformation adapting it to the latter.

To demonstrate the method, we calculated thermal decay rates of a particle placed at the metastable ground state of a cubic potential. On the one hand the decay rate was obtained by the Hamilton function of a point particle using the classical and quantum normal forms. On the other hand this was done by the presented method with a frozen Gaussian wave function of different width described in the framework of a variational approach to the Schr\"odinger equation and the results show an excellent agreement with the well established methods.

The application of the variational approach to wave packets within the framework of a time-dependent variational principle together with the demonstrated mapping to classical phase space further allows to calculate decay rates for quantum mechanical systems where an analogous classical Hamiltonian is not directly accessible. 

Such systems are, e.g.\ Bose-Einstein condensates with additional long-range interaction which have been successfully approached variationally with coupled Gaussian wave functions \cite{Rau2010a, Rau2010b}. In case of an attractive short-range contact interaction, their ground state is metastable and a sufficient thermal excitation may lead to the collapse of the condensate by crossing a barrier which is formed by an unstable excited state that reveals the properties of a saddle-centre-\ldots-centre equilibrium point. Thus, the variational approach to BECs is predestined for the procedure presented in this paper and we refer the reader to reference \cite{Junginger2011c} for the application to these ultra-cold gases.

Of course, this method can also be applied to the field of chemical reactions where the behaviour of reactants can be studied without approximations such as in the framework of an ab initio calculated Born-Oppenheimer potential energy surface if a suitable parametrization of the system's wave function has been found.

\section*{Acknowledgement}

This work was supported by Deutsche Forschungsgemeinschaft. A.\,J. is grateful for support from the Landesgraduiertenf\"orderung of the Land Baden-W\"urttemberg. We also thank Thomas Bartsch for fruitful discussions.

\section*{Appendix: Example}

In order to give a descriptive illustration of the variational parameters' mapping to classical phase space, we present in this appendix a numerical example for the parameters $a=1.9$, $\alpha=10$ and $\gamma=2$ in fifth order of the equations of motion and sixth order of the energy functional, respectively.

For this set of parameters the Taylor expansion \eref{eq-DGL-series} of the equations of motion \eref{eq-zdot-model-potential-a}--\eref{eq-zdot-model-potential-b} in the vicinity of the stable fixed point reads
\begin{eqnarray}
	\dot{x}_1 	&=	3.8 x_2, 	
	\label{appendix-expansion-DGL-a}\\
	\dot{x}_2 	&=	-3.67852 x_1 + 0.519391 x_1^2 
	\label{appendix-expansion-DGL-b}%
\end{eqnarray}
and for the corresponding expansion \eref{eq-energy-expansion} of the energy functional \eref{eq-energy-functional} one obtains
\begin{equation}
	E = 1.90363 + 0.484015 x_1^2  - 0.0455606 x_1^3 + 0.5 x_2^2 .
	\label{appendix-expansion-E}
\end{equation}
At this point, there is no obvious way to derive equations \eref{appendix-expansion-DGL-a} and \eref{appendix-expansion-DGL-b} from equation \eref{appendix-expansion-E} via Hamilton's equations.

With the eigenvalues $\ii\omega_{1,2}=\pm 3.73877\ii$, the normal form expansion of the equations of motion and the introduction of the canonical variables $\tilde{q}=x_1$, $\tilde{p}=x_2$ results in 
\begin{eqnarray}
	\dot{\tilde{q}} 	&=	~~\, 3.73877 \ii \tilde{q} - 0.063123 \ii \tilde{p} \tilde{q}^2 - 0.00250446 \ii \tilde{p}^2 \tilde{q}^3,	\\
	\dot{\tilde{p}} 	&=	- 3.73877 \ii \tilde{p} + 0.063123 \ii \tilde{p}^2 \tilde{q} + 0.00250446 \ii \tilde{p}^3 \tilde{q}^2
	\label{appendix-after-NFT-DGL}%
\end{eqnarray}
for equations \eref{eq-structure-monomials-qp-a} and \eref{eq-structure-monomials-qp-b}, where the higher-order terms have arisen during the transformations. The coefficients on the right-hand side of these two equations only differ in sign and therefore fulfil the conditions of integrability \eref{eq-integration-conditions-a}--\eref{eq-integration-conditions-c}. Consequently, they can easily be integrated to a Hamilton function
\begin{equation}
	\Htilde = \Htilde_0 + 3.73877 \ii \tilde{p}\tilde{q}  - 0.0315615 \ii (\tilde{p}\tilde{q})^2 - 0.0008482 \ii (\tilde{p}\tilde{q})^3 . 
	\label{appendix-Htilde}
\end{equation}

However, after having applied the change of coordinates corresponding to the transformation above, the energy functional \eref{eq-energy-series-J} reads
\begin{equation}
	E = 1.90363 + 0.983756 \tilde{p}\tilde{q} - 0.00830456 (\tilde{p}\tilde{q})^2 - 0.000219661 (\tilde{p}\tilde{q})^3
		\label{appendix-after-NFT-E}
\end{equation}
whose coefficients do not agree with equation \eref{appendix-Htilde}.

To achieve this, we apply the scaling of the phase space variables $\tilde{\qq}\to\qq$ and $\tilde{\pp}\to\pp$ according to equations \eref{eq-nu-scaling}--\eref{eq-mu-power-series} which after integration of the equations of motion together with the definition of the action-variable $J= \ii pq$ leads to
\begin{eqnarray}
	\Htilde =&~ \Htilde_0 - 3.73877 J + 0.0315615 \ii \mu_1 J^2 \nonumber \\
			& - (0.000834821 \mu_1^2 + 0.021041 \mu_2 ) J^3, \\[0.5em]
	E =&~	1.90363 + 0.983756 \ii \mu_1 J + ( 0.00830456 \mu_1^2 - 0.983756 \mu_2) J^2  \nonumber \\
			& + ( 0.000219661\ii \mu_1^3 + 0.0166091\ii \mu_1 \mu_2 ) J^3.
	\label{appendix-after-mu-scaling}%
\end{eqnarray}
Setting the constant of integration to $\Htilde_0=1.90363 $ and choosing $\mu_1 = 3.8005 \ii$, $\mu_2 = 0.0$  finally yields
\begin{eqnarray}
	H(J) &= \Htilde (J) = E(J)  \nonumber \\
	&= ~1.90363 - 3.73877 J  - 0.119949 J^2 + 0.012058 J^3 
\end{eqnarray}
which is the desired Hamilton function in third order of the action-variables.

We note that the vanishing scaling parameter $\mu_2 = 0.0$ is a consequence of the ansatz \eref{eq-frozen-gaussian} where the real and imaginary part of the variational parameter $z$ already have the meaning of position and momentum of the wave function. In general, all the $\mu_{\mm}$ will be non-zero.

\section*{References}

\begin{thebibliography}{10}

\bibitem{Jaffe1999}
C.~Jaff\'e, D.~Farrelly, and T.~Uzer.
\newblock Transition state theory without time-reversal symmetry: Chaotic
  ionization of the hydrogen atom.
\newblock {\em Phys. Rev. Lett.}, 84(4):610--613, 2000.

\bibitem{Eckhardt1995}
{B. Eckhardt}.
\newblock Transition state theory for ballistic electrons.
\newblock {\em J. Phys. A: Math. Gen.}, 28(3469), 1995.

\bibitem{Komatsuzaki1999}
{T. Komatsuzaki} and {R. S. Berry}.
\newblock Regularity in chaotic reaction paths: {I}. {A}r{$_6$}.
\newblock {\em J. Chem. Phys.}, 110(9160):3, 1999.

\bibitem{Oliveira2002}
H.~P. de~Oliveira, A.~M. Ozorio~de Almeida, I.~Dami\~ao Soares, and E.~V.
  Tonini.
\newblock Homoclinic chaos in the dynamics of a general bianchi type-ix model.
\newblock {\em Phys. Rev. D}, 65(8):083511, 2002.

\bibitem{Garrett1983}
{B. C. Garrett}, {D. G. Truhlar}, {A. F. Wagner}, and {T. H. Dunning}.
\newblock Variational transition state theory and tunneling for a
  heavy-light-heavy reaction using an ab initio potential energy surface.
  $^{37}${C}l+{H}({D})$^{35}${C}l$\to${H}({D})$^{37}${C}l+$^{35}${C}l.
\newblock {\em J. Chem. Phys.}, 78(4400), 1983.

\bibitem{Aoiz1994}
{F. J. Aoiz}, {L. Bafiares}, {V. J. Herrero}, {V. Shez Rhbanos}, {K. Stark},
  and {H.-J. Werner}.
\newblock Classical dynamics for the {F} + {H}$_2$ $\to$ {HF} + {H} reaction on
  a new ab initio potential energy surface. {A} direct comparison with
  experiment.
\newblock {\em Chemical Physics Letters}, 223:215--226, 1994.

\bibitem{Truong1990}
{T. N. Truong} and {D. G. Truhlar}.
\newblock Ab initio transition state theory calculations of the reaction rate
  for {OH}+{CH}$_4 \to$ {H}$_2${O}+{CH}$_3$.
\newblock {\em J. Chem. Phys}, 93(1761), 1990.

\bibitem{Soto1992}
{M. R. Soto} and {M. Page}.
\newblock Ab initio variational transition-state-theory reaction-rate
  calculations for the gas-phase reaction {H}+{HNO} $\to$ {H}$_2$+{NO}.
\newblock {\em Journal of Chemical Physics}, 97(7287), 1992.

\bibitem{Junginger2011c}
{A. Junginger}, {M. Dorwarth}, {J. Main}, and {G. Wunner}.
\newblock Transition state theory for wave packet dynamics. {II}. {T}hermal
  decay of {B}ose-{E}instein condensates with long-range interaction.
\newblock following paper, submitted to J. Phys. A, 2011.

\bibitem{Rau2010a}
S.~Rau, J.~Main, and G.~Wunner.
\newblock Variational methods with coupled gaussian functions for
  {B}ose-{E}instein condensates with long-range interactions. {I}. {G}eneral
  concept.
\newblock {\em Phys. Rev. A}, 82:023610, 2010.

\bibitem{Rau2010b}
S.~Rau, J.~Main, H.~Cartarius, P.~K\"oberle, and G.~Wunner.
\newblock Variational methods with coupled gaussian functions for
  {B}ose-{E}instein condensates with long-range interactions. {II}.
  {A}pplications.
\newblock {\em Phys. Rev. A}, 82:023611, 2010.

\bibitem{Waalkens2008}
H.~Waalkens, R.~Schubert, and S.~Wiggins.
\newblock Wigner's dynamical transition state theory in phase space: classical
  and quantum.
\newblock {\em Nonlinearity}, 21(1):R1, 2008.

\bibitem{Murdock2010}
{J. Murdock}.
\newblock {\em Normal {F}orms and {U}nfoldings for {L}ocal {D}ynamical
  {S}ystems}.
\newblock Springer, 2010.

\bibitem{Schubert2006}
R.~Schubert, H.~Waalkens, and S.~Wiggins.
\newblock Efficient computation of transition state resonances and reaction
  rates from a quantum normal form.
\newblock {\em Phys. Rev. Lett.}, 96(21):218302, 2006.

\bibitem{McLachlan1964}
{A. D. McLachlan}.
\newblock A variational solution of the time-dependent {S}chr{\"o}dinger
  equation.
\newblock {\em Molecular Physics}, 8(1):39--44, 1964.

\bibitem{Fabcic2008}
{T. Fab\v{c}i\v{c}}, {J. Main}, and {G. Wunner}.
\newblock Time propagation of constrained coupled {G}aussian wave packets.
\newblock {\em J. Chem. Phys.}, 128(044116), 2008.

\bibitem{Waalkens2004}
{H. Waalkens} and {S. Wiggins}.
\newblock Direct construction of a dividing surface of minimal flux for
  multi-degree-of-freedom systems that cannot be recrossed.
\newblock {\em J. Phys. A: Math. Gen.}, 37:L435--L445, 2004.

\bibitem{MacKay1990}
R.~S. Mackay.
\newblock Flux over a saddle.
\newblock {\em Physics Letters A}, 145(8-9):425 -- 427, 1990.

\bibitem{Toller1985}
M.~Toller, G.~Jacucci, G.~DeLorenzi, and C.~P. Flynn.
\newblock Theory of classical diffusion jumps in solids.
\newblock {\em Phys. Rev. B}, 32(4):2082--2095, 1985.

\end{thebibliography}

\end{document}